\documentclass[conference]{IEEEtran}
\IEEEoverridecommandlockouts
\pdfoutput=1
\usepackage{cite}
\usepackage{amsmath,amssymb,amsfonts}
\usepackage{algorithmic}
\usepackage{graphicx}
\usepackage{textcomp}
\usepackage{wrapfig}
\usepackage[framemethod=tikz]{mdframed}
\usepackage{tcolorbox}
\usepackage{lipsum}
\usepackage{multicol}

\newenvironment{WrapText}
{\begin{figure*}\tcolorbox\begin{multicols}{2}}
{\end{multicols}\endtcolorbox\end{figure*}}

\usepackage{color}    
\def\BibTeX{{\rm B\kern-.05em{\sc i\kern-.025em b}\kern-.08em
    T\kern-.1667em\lower.7ex\hbox{E}\kern-.125emX}}

\begin{document}

\title{Rethinking Blockchain Security: Position Paper (preprint)
\thanks{This work was supported in part by the National Research Foundation (NRF), Prime Minister's Office, Singapore, under its National Cybersecurity R\&D Programme (Award No. NRF2016NCR-NCR002-028) and administered by the National Cybersecurity R\&D Directorate.}
}

\newcommand{\affa}{$^*$}
\newcommand{\affb}{$^\dagger$}

\author{
\IEEEauthorblockN{Vincent Chia\affa, Pieter Hartel\affb, Qingze Hum\affb, Sebastian Ma\affa, Georgios Piliouras\affb\ \\ Dani\"el Reijsbergen\affb, Mark van Staalduinen\affa, Pawel Szalachowski\affb }
\IEEEauthorblockA{\\[-0.2cm]\affa The Netherlands Organisation for Applied Scientific Research (TNO) \\ \textit{\{vincent.chia, sebastian.ma, mark.vanstaalduinen\}@tno.nl} \\[0.1cm]
\affb Singapore University of Technology and Design \\ \textit{\{pieter\_hartel, georgios, daniel\_reijsbergen, pawel\}@sutd.edu.sg} \\ \textit{qingze\_hum@mymail.sutd.edu.sg}}
}

\maketitle
\begin{abstract}
Blockchain technology has become almost as famous for incidents involving security breaches as for its innovative potential. 
We shed light on the prevalence and nature of these incidents through a database structured using the STIX format. Apart from OPSEC-related incidents, we find that  the nature of many incidents is specific to blockchain technology. Two categories stand out: smart contracts, and techno-economic protocol incentives. For smart contracts, we propose to use recent advances in software testing to find flaws before deployment. For protocols, we propose the PRESTO framework that allows us to compare different protocols within a five-dimensional framework.

\end{abstract}

\begin{IEEEkeywords}
blockchain, security, smart contracts
\end{IEEEkeywords}

\section{Introduction}
\label{sec:introduction}
The idea of the blockchain --- an open ledger maintained and extended by a peer-to-peer network --- is shaping up to become a foundational technology underpinning not just cryptocurrencies but also a diverse range of other applications \cite{iansiti2017truth,underwood2016blockchain,swan2015blockchain,tapscott2016blockchain}. However, its novelty means that its surrounding security infrastructure has some way to go to reach the level of more established digital services such as online banking. Combined with the  increased valuation of tokens, this has made cryptocurrency platforms a popular target for attacks. 
This is witnessed by several recent high-profile thefts or other losses of cryptocurrency tokens worth huge sums of money. These include the collapse of Mt.Gox (\$460 million stolen \cite{mcmillan2014inside}), the Coincheck hack (\$400 million stolen \cite{menegus2018coincheck}), the DAO hack (\$50 million stolen \cite{siegel2016understanding}), and the Parity bug (\$160 million frozen \cite{oleary2017parity}). The first two of these incidents involve `traditional' OPSEC ---  trust, key, and information management issues that are no different for a cryptocurrency exchange than for a bank. However, the latter two incidents are of a different nature as the flaws themselves are \emph{on the blockchain}. On one hand this is good for transparency, because the exact nature of the bug is publicly visible and the movement of stolen funds can be tracked. However, because blockchains are append-only, it is impossible to fix security flaws or recover lost funds without a hard fork. A similar point can be made for unintended consequences of the crypto-economic incentives that are inherent to blockchain consensus protocols. Examples include the emergence of mining pools, the potential of 51\% attacks, or block withholding attacks against the entire network \cite{eyal2014majority} or the incentive structure within pools.\footnote{{https://www.reddit.com/r/Bitcoin/comments/28242v/eligius\_falls\_victim \_to\_blocksolution\_withholding/}} Since solutions require profound changes to the protocol, hard forks again tend to become a necessity.

In this paper, we present an overview of the security challenges that are unique to blockchain technology, and argue that these are best confronted through extensive testing before deployment. The structure of the paper is as follows. In Section~\ref{sec:incident_database} we present a database of blockchain incidents that was compiled in a structured manner. From this database, we learn that while traditional OPSEC-related issues remain the largest source of incidents, a considerable proportion (roughly one third) of the incidents are blockchain-specific. These can further be subdivided into two categories: smart contracts and protocol incentives. These are discussed in Section~\ref{sec:smart_contract_security} and Section~\ref{sec:consensus_protocol} respectively. Section~\ref{sec:conclusions} concludes the paper.

\section{Blockchain Incident Database}
\label{sec:incident_database}

As mentioned in the introduction, the increased market capitalisation of cryptocurrencies, combined with the lack of maturity of the surrounding security ecosystem, has led to a considerable number of security incidents.
Blockchain incidents are defined as attacks resulting in a loss of assets.
These incidents have been reported in different sources and with different levels of detail.

In this section, we present a first structured attempt to collect and curate a database of blockchain incidents.
The resulting database supports analytics that we believe will improve the cybersecurity of blockchain applications.
With the blockchain incident database, the different risks can be identified that should at least be known to the blockchain community.
The incident database can serve as a reference for blockchain developers to improve the security of their solutions, and it will also improve the understanding of known vulnerabilities in blockchain deployments.
Incident types span from traditional cyber attacks on cryptocurrency exchanges such as Mt. Gox~\cite{mcmillan2014inside}, to exploits of vulnerable smart contracts~\cite{siegel2016understanding} and hacked computers that mine cryptocurrencies~\cite{mak2018how}.

\subsection{Database Creation}
Our goal is to create an easy-to-use database containing a representative set of blockchain incidents.
The collection of the incidents consists of four steps: identification, description, classification, and review of incidents.

\subsubsection*{Identification}
We assume that all relevant incidents can be identified in open sources~\cite{Eeten2011}.
First, there are specific sources, such as the ``Blockchain Graveyard''\footnote{{https://magoo.github.io/Blockchain-Graveyard/}} and the Ethereum Blog Security Archives.
Additionally, Google searches for specific terms, such as ``hacking for Bitcoin mining'' and ``conducted 51\% attacks'' provide further data.
These sample queries are inspired by the security challenges identified by ENISA~\cite{enisa2017distributed}.
Given that this is a first attempt to build the database, we do not claim completeness.
However, our ultimate goal is to develop a representative dataset through crowdsourcing.

\subsubsection*{Description}
For each incident we store a description with ample references, such as screenshots, and web archives of original data for further verification.
For the incident descriptions, we use STIX~\cite{barnum2012standardizing}, which is an international standard for sharing cybersecurity incident data.
We use 14 standard STIX objects as our fields, extended with a new field --- `Loss Estimation (crypto)' --- describing the cryptocurrency involved, such as Bitcoin (BTC), Ether (ETH) or any altcoin lost in the incident.

The incidents can be viewed and edited using a database platform, which is running in the cloud and provided as a web service.
Moderated registration on the platform will soon open.

Incidents are described with as many references as possible to give a view of the incident that is as broad as possible.
Nevertheless, open sources are limited and we cannot guarantee the integrity of the complete picture.
We aim to ensure verifiability of incidents by third parties.
However, as most companies do not fully disclose information, we acknowledge that some information might be missing.
Therefore, each incident is editable in order to add the latest insights.

\subsubsection*{Classification}
Based on analysis of the first version of the incident database with 86 incidents, we propose three main classes of incidents:

\hspace{0.4cm}{\bf OPSEC:}
Incidents compromising an organisation or individual's control of information and access to business-critical assets.

\hspace{0.4cm}{\bf Smart Contracts:}
Incidents resulting from improperly written smart contracts deployed and executed on a blockchain.

\hspace{0.4cm}{\bf Consensus Protocol Incentives:}
Incidents arising from malicious exploitation of consensus protocols that create opportunities and benefits for blockchain participants.

OPSEC is a container class of blockchain incidents based on traditional cyber attacks~\cite{kpmg2017securing}.
Smart Contract breaches are further described in Section~\ref{sec:smart_contract_security}, whereas Consensus Protocol Incentive vulnerabilities are discussed in Section~\ref{sec:consensus_protocol}.

\subsubsection*{Review}
The final step in the process of building the database is the review, which ensures that independent parties verify incident data.
Thus far we have employed students to provide a somewhat independent view on the incidents.
Ultimately, we hope that we will be able to incentivise the blockchain community to contribute to the review process.

\subsection{Incident Analytics}
Most of the added value of the incident database is provided by the analytics, such as the time to incident discovery and the monetary losses incurred.
For example, even for our relatively small collection of incidents covering the period 2011-2018, we estimate that some US\$ 3.55B has been lost.

We propose to build a number of more advanced analytic capabilities as follows:
\begin{itemize}
\item Develop a deep understanding of the attack vectors within the different classes of incidents.
\item Analyse the time from attack to detection, to see how cybersecurity changes as a result of blockchain technology evolution.
\item Investigate the correlation between specific blockchain technologies and classes of incidents.
\item Create a timeline of the losses based on incidents.
\item Discover trends in the number of incidents and losses over time, to see whether countermeasures are effective.
\item Map geographical origins of attacks.
\end{itemize}

Based on our first collection of blockchain incidents we can draw the following preliminary conclusions.

\hspace{0.4cm}{\bf OPSEC:}
The majority of incidents in our current database occurred due to a lack of sufficient OPSEC measures (about 66\%).
The key blockchain component is the opportunity for attackers to confiscate enormous amounts of cryptocurrencies, resulting in incidents with millions of lost assets.
To prevent the OPSEC class of incidents, standard cybersecurity solutions are available.
For this reason, our position is to consider OPSEC out-of-scope regarding research to new solutions.

\hspace{0.4cm}{\bf Smart Contracts:}
This category makes up about 22\% of our incident database.
These incidents occur when a smart contract does not work the way it was intended.
The transparency of blockchains makes it possible to audit all published smart contracts, which can be supported by tools.
For this reason, our position is to conduct research into new solutions focused on smart contract testing, as we discuss in Section~\ref{sec:smart_contract_security}.

\hspace{0.4cm}{\bf Consensus Protocol Incentives:}
Consensus protocol attacks are more difficult to detect than smart contract related attacks, as the effect is usually the improper mining of a block or the censorship of nodes.
It is hard for miners to mitigate or even detect attacks.
Furthermore, the majority of the incidents arising from incentives are due to second-order, unintended effects of blockchain use. 
These incidents make up roughly 12\% of our database.
Our position is to use a formal framework --- see Section~\ref{sec:consensus_protocol} --- to evaluate the incentives of the consensus protocols in advance, instead of realising the criminal opportunities after an attack.

\section{Smart Contract Security}
\label{sec:smart_contract_security}
The concept of smart contracts is a relatively new paradigm that enables
different mutually untrusting parties to encode and enforce their agreements.
Practical implementations of smart contracts emerged with the advent of
blockchain platforms based on distributed consensus protocols.  Currently,
mainstream smart contract platforms, like Ethereum~\cite{buterin2014next},
follow the model of replicated execution. In this model, the code of all smart
contracts, and all calls to contract methods, are irreversibly appended to a
blockchain. Every node participating in the blockchain protocol reads code and
calls from the blockchain, instantiates the same code, and executes the same
calls, getting the same results and maintaining the same state.

From the security point of view, this model has many important implications.
First of all, the development life cycle of smart contracts is significantly
different from the traditional software development life cycle, where testing,
integration, and maintenance are repeatable.  Since a smart contract's code is unchangeable after being appended
to a blockchain, developers have to bind code to a variable
if they wish to modify the behaviour of their contracts later on. In that context, the development life cycle of smart contracts is much
different from standard software that can be patched and fixed on-the-fly. One could
even argue that as code changes are impossible, the development of smart
contract is not a cycle anymore.

Subsequently, mistakes in code (like logical errors or even typos) or mistakes
in the use of smart contracts (like calling a method with wrong parameters) are
irreversible and costly.
This argument is supported by
our analysis of a sample contract as shown in Case Study~I.
In fact, hackers
are constantly looking for vulnerabilities in deployed smart contracts, treating
them as specific bug-bounty programs~\cite{kuehn2014analyzing}, where they can
\textit{get paid} for exploiting vulnerabilities.

Unfortunately, it is difficult to prevent such attacks as deployed code cannot
be patched retrospectively. Instead, when a vulnerability is detected, a new smart
contract has to be deployed to fix it.  Although it is difficult for developers,
the immutability of smart contracts is seen by the community as something
positive.  Consider, for example, a case where a developer could fix a bug in an already
deployed smart contract.  Such a change could cause collateral damage as other
smart contracts might rely on the \textit{fixed} functionality of the changed
contract.  Furthermore, what the developer of the smart contract considers a
bug could be a property for developers of relying smart contracts.  In that
sense, smart contracts are similar to legislation (just to recall the sentence
``Code is Law''~\cite{lessig1999code}), and therefore should be stable and
carefully prepared before deployment.

Our observations lead us to the conclusion that verification and testing are
especially important in smart contract development, and should be an integral part of
the analysis and design steps (and not only follow the implementation step as in the
traditional development life cycle).

\subsection{Testing and Verification}
\label{sec:smart:test}
The aim of security testing is to determine whether a program has a vulnerability
that can be exploited by an attacker~\cite{Potter2004}. Smart contracts are programs
and may therefore contain vulnerabilities~\cite{Everts2018}. Smart contracts are
usually short but they are also concurrent~\cite{Sergey2017} and permanent, so
probably more difficult to get right than ordinary programs~\cite{Delmolino2016}.

There is a relatively recent review of smart contract vulnerabilities~\cite{Atzei2017},
which shows that smart contracts have classical issues (like off-by-one errors)
as well as specific issues (like running out of gas).  Smart contracts should
therefore be subjected to security analysis and testing.

There are several types of testing and analysis tools that can be used on smart
contracts, such as running hand crafted tests on a fast test network
with the truffle framework\footnote{See {http://truffleframework.com}},
fuzzing of the input of the contract, mutating of the code of the
contract~\cite{Jia2011}, static analysis of properties of the contract~\cite{Luu2016},
model checking of behaviours of a model of the contract~\cite{Chaudhary2015}, and
theorem proving of properties of the program~\cite{Hira2017}. There are also runtime
verification techniques, such as proof carrying code~\cite{Holthusen2016}.

Some tools require test engineers to be highly skilled. For example the use of
theorem provers requires considerable skill, whereas automatic static analysis
tools\footnote{See for example {https://securify.ch}} are simple to use. Ease of use
does not necessarily mean that the results are uninteresting; this depends on
the sophistication of the tool.

We believe that there is enough low-hanging fruit for us to be able to afford focussing on
tools that are easy to use, such as mutation and fuzzing tools.  We are unaware
of existing literature on the application of fuzzing and mutating to smart contracts.
These techniques can form the basis of a powerful test suite --- i.e., one that has good code coverage and helps the test engineer to distinguish successful
traces from failures.


Writing tests is both tedious and error prone. Fuzzing varies the inputs to a
program under test, which helps to extend the test suite. Mutating varies the
code of a program under test, which helps to increase the test coverage.
However, even with these tools significant creativity is needed on the part of
the test engineer to develop a comprehensive test suite for a program or smart
contract.

We list four approaches that we believe can help the test engineer. These are all
low-risk high-return approaches, building on a wealth of related work.  Most
approaches regard a contract from a different point of view, which gives the test
engineer insights that he/she might not otherwise have acquired. For example, if
a lottery contract has never paid out, a security problem is likely.

\begin{WrapText}
\subsection*{Case Study I: Vitaluck is a poorly tested DApp}
\label{subsec:case_study_vitaluck}
Vitaluck was a DApp launched on 26 Jan 2018 by a developer known
on reddit as luckyico. This DApp caught our attention because of
the name.\footnote{See
https://coinsnews.com/very-excited-to-share-my-first-solidity-project-vitaluck}
Vitaluck is a raffle game with the following advertised rules:\footnote{See
{https://www.vitaluck.com/my-bets/}}
\begin{itemize}
\item Each play generates a random number between 0~and~1000. If
      the player's number is bigger than the previous number you
      win the Jackpot.
\item For every bet, the commission is 10\%.
\item For every losing bet, 10\% is paid to the previous winning
      address, and 80\% goes in the Jackpot.
\item The minimum bet is 0.05 ETH.
\item When the winning number exceeds 900, it is reset to~1.
\end{itemize}

We conducted an audit of the source of the contract, and its history on the
blockchain.\footnote{See 
{https://etherscan.io/address/0xef7c7254c290df3d167182356255cdfd8d3b400b}}
We wrote one script to analyse the transactions of the contract.
We also hand-translated the contract into an equivalent JavaScript
program so that we could simulate the game for arbitrary numbers
of rounds and players. The game has been played 27 times by 10
accounts, at least one of which is owned by the author of the game.

We discovered six typical smart contract security issues:
\begin{itemize}
\item The contract uses the time stamp of the last block on the
      blockchain as a source of low-entropy randomness. This means
      that the ticket numbers can be predicted, which makes it easy
      to cheat.
\item The reset method subtracts 0.05 ETH from the jackpot without
      checking whether there is enough ETH in the jackpot.
\item The jackpot can never be completely emptied, so the residual
      funds cannot be claimed.
\item The list of previous bets is stored but never pruned, so the
      list grows indefinitely.
\item The rules suggest that the commission is 10\%, but the
      commission is actually higher. The commission slowly decreases,
      but after many thousands of rounds it is still higher than
      10\%.
\item There is no exception handling in the contract, whereas it
      uses methods like transfer that can throw exceptions.
\end{itemize}

By way of responsible disclosure, we tried to contact the author
of the Vitaluck contract in a variety of ways to discuss our findings
but he/she has not replied. Since the DApp has been disabled, we
believe that publication of this case study is justified, as it
will serve as a warning to the community not to skip testing too
lightly.



According to the post on coinsnews cited above, Vitaluck was the
first DApp built by the author. This would go someway towards
explaining why there are so many issues in a relatively short and
straightforward contract.  On the other hand, if we could find six
issues in about a day's work, the author should have been able to
find them also.  We believe that the contract simply has not been
tested well, which underlines the position taken in this paper.
\end{WrapText}

\begin{WrapText}
\subsection*{Case Study II: Solidity compiler bugs introduce vulnerabilities in high-value contracts}
\label{subsec:case_study_compiler_bugs}

We analysed the top 1,000 contracts by balance of the almost 20,000
verified smart contracts on etherscan.io. A contract is verified
if its source code and compiler version are made public.\footnote[4]{See
 {https://etherscan.io}}  Making this information public
improves transparency and should therefore increase the confidence
in the contract.

\begin{wraptable}[14]{r}{0.3\textwidth}
\vskip-0.4cm
\begin{tabular}{r|lr} 
$n$ & version         & release date \\ \hline
     120     & v0.1-3.*        & 2015 \\
      88     & v0.4.0-9        & 2016 \\
      14     & v0.4.10         & Mar 2017 \\
      68     & v0.4.11         & May 2017 \\
      41     & v0.4.12-14      & Jul 2017 \\
     122     & v0.4.15-16      & Aug 2017 \\
      36     & v0.4.17         & Sep 2017 \\
     187     & v0.4.18         & Oct 2017 \\
     207     & v0.4.19         & Nov 2017 \\
      78     & v0.4.20         & Feb 2018 \\
      39     & v0.4.21         & Mar 2018  \\[0.2cm]
    1000     & \multicolumn{2}{l}{total}
 \end{tabular}
 \end{wraptable}

The contract with the largest balance holds 1,500,000 ETH, the
smallest balance is just 0.9 ETH.  The sum total of the balance of
the 1,000 contracts is just over 5 million ETH, or about 5\% of the total
amount of ETH tokens in circulation.  Therefore we believe our selection
of contracts to be representative for contracts on Ethereum.

Some of the 1,000 verified contracts were compiled by a version of
the Solidity compiler that is more than a year old. Older versions
have known bugs that introduce vulnerabilities in the compiled code.
The table to the left lists how many contracts ($n$) have been compiled with
the various compiler versions.

Of the 1,000 verified high-value contracts, almost half has a
vulnerability caused by a known bug in the Solidity compiler. The following table lists how many contracts have a low/medium/high severity issue:

\begin{wraptable}{r}{0.26\textwidth}
\begin{tabular}{r|l} 
\#contracts  & severity level \\ \hline
       128   & high \\
       192   & medium \\
        10   & low \\
       158   & very low \\[0.2cm]
       488   & total
 \end{tabular}
 \end{wraptable} 

About one third of the verified high-value contracts has a medium
to high severity vulnerability. Since a contract is baked into the
blockchain, it cannot simply be recompiled with an up-to-date
version of the compiler to remove the vulnerabilities. This is
clearly an issue of smart contracts that should be resolved. Again
this case study supports our position that more testing is needed.
\end{WrapText}

\subsubsection{Documentation for developers and testers}
The first approach is to improve the documentation on smart contract development
and testing. It is difficult to develop smart contracts because the
documentation is scant and poorly organised. The Ethereum yellow paper~\cite{Wood2017} can be
hard to read,
the official documentation of the Solidity language is primarily a collection of examples~\cite{Ethereum2018},
there are many resources that tell parts of the story and
invariably the developer ends up searching the web\footnote{See for example {https://stackoverflow.com}} for
questions that might be related to the issues, hoping to find usable answers.

Consider as an example how to determine whether a transaction has run out of
gas. The Byzantium version of Ethereum returns a status flag in the transaction
receipt to indicate this. Other versions require the programmer to compare the
actual amount of gas used to the maximum amount of gas passed to the
transaction. If the two are equal, the probability is high that indeed the
transaction has run out of gas. The probably is not 100\%, since it is
theoretically possible that the transaction used exactly the maximum amount of
gas. If the developer knows what he/she is looking for, the information can be
found. But many unsuspecting developers will probably make mistakes because
they cannot find the information they are looking for.

We believe that a handbook for smart contract developers and testers is long
overdue.

\subsubsection{Fuzz the inputs to smart contracts}
The second approach is to fuzz the inputs of the smart contracts.  Fuzzing has
been proved as an effective and fast technique for finding security
vulnerabilities~\cite{sutton2007fuzzing,godefroid2012sage}.  It can be applied
directly to a smart contract by applying invalid and unexpected inputs and
observing behaviour of the smart contract.  Fuzzing can also be used for finding
bugs specific to smart contracts.  For example, by varying the limit on the
available amount of gas, transactions can be aborted at arbitrary points,
possibly leaving the contract in an insecure state.  Another example is to
randomise the order of calls, as in some cases a miner or underlying network can
influence this, and observe the impact.

\subsubsection{Mutate the code of smart contracts}
The third approach is to develop a mutation tool for smart contracts. This has
been proposed
before,\footnote{See {https://github.com/ethereum/solidity/issues/1172}} but the
attempt has been abandoned. The challenge is for the mutation generator to
understand enough of the semantics of the smart contract language to generate
only useful mutants.  One possible route is to export the abstract syntax tree
from the smart contract compiler, mutate it, and import the mutant again in the
compiler.

\subsubsection{Search the blockchain for tests}
The fourth approach to address the problem of insufficient tests is to search
the blockchain for traces of already deployed smart contracts. The blockchain
contains all code and data needed to recreate the state of the world at each
time a block was mined. Therefore, the traces from the blockchain contain
enough information to generate tests. This idea is similar to using the counter
examples produced by model checkers for testing purposes~\cite{Fraser2009}. The
difference is that we do not have to create a model of the contract; instead the
trace originates from the contract itself. Both testnet and livenet deployments
of smart contracts can be searched for traces. These deployments are probably
not fully tested; hence the test engineer will have to inspect the generated
tests carefully. We believe that such generated tests could complement a test
suite written by hand.

Most of the techniques that we describe in this chapter exist
already. However, the Dapps that we have analysed show that not all
developers master the relevant techniques. So there is a need to
make these more accessible, starting by improved documention.

\subsection{Secure Smart Contracts Distribution}
Since smart contract platforms have associated cryptocurrencies, they are
attractive medium for fraud, often implemented as
smart contracts.  In traditional software distribution there are many models to
handle such problems.  For instance, open-source projects are driven by
communities who have interest in fixing bugs and increasing the quality of these
projects.  Similarly, centralised software repositories (like Apple App Store,
Google Play, or Ubuntu repositories) are controlled by companies who have
incentives in removing malicious code, as it directly affects their clients.

By contrast, the distribution of smart contracts does not follow any of these
models.  Smart contracts are submitted by developers (anyone can act as a
developer) and are appended to a blockchain by miners (i.e., nodes that
participate in the underlying consensus protocol).  The irreversibility and
censorship-resistance of smart contract platforms do not allow anyone to remove
an added smart contract (even if it evident that it is malicious), so providing
security in such an architecture is a challenging task.

In order to improve on this, we propose a proactive smart contract publishing
architecture. The main observation is that miners of smart contract platforms
have conflicting incentives:
\begin{itemize}
    \item they have incentive to add every smart contract submitted (even
        malicious), as appending code to the blockchain rewards them, and
    \item they have incentive to keep the ecosystem secure, as otherwise it
        might lose popularity affecting their rewards.
\end{itemize}

A high-level idea of our architecture is to keep adding all submitted smart
contracts to the blockchain, and to keep testing their security properties.
Then for every found (potential) vulnerability a warning can be signalled in the blockchain, such that everyone can see it.
In particular we distinguish two deployment models of such an architecture.

In the first model, a miner before adding a smart contract tests it with various
testing methodologies and tools (see 
Section~\ref{sec:smart:test}). 
The miner publishes the
smart contract along with the obtained testing results. For instance, these
results can indicate that the smart contract has a potential backdoor or a kill
switch, or it has some features of a Ponzi
scheme~\cite{bartoletti2017dissecting}.  When appended to the blockchain these
results are immutable and visible to everyone, thus can act as reliable
warnings.  An alternative deployment model, is to introduce a dedicated service
which would constantly test existing and incoming smart contracts, and would
publish testing outcomes (i.e., warning) at a pre-defined location.  An advantage of
this approach, is that such a service can be upgraded to include new test methods and
vectors, and does not depend on miners who might be be unwilling to conduct
additional work~\cite{luu2015demystifying}. 
We leave details of this architecture as a future work, however, we envision
that such a service can be maintained by community contributors and its security
can be enhanced by means like a trusted execution environment.


\section{Cryptoeconomic Protocols \& the PRESTO framework}
\label{sec:consensus_protocol}
At the core of blockchain platforms such as Bitcoin and Ethereum lies the emergence of new protocols which aim to align the interests of several self-interested parties that do not necessarily trust each other. Different platforms combine different ideas and explore different trade-offs. These ideas can be better understood as points in a high-dimensional space, and the goal of a protocol designer is to develop protocols that lie on the Pareto frontier of the design space, i.e., systems whose desirable properties cannot all be simultaneously improved.  

An important tool towards understanding the design space better is the organisation of ``desirable" properties onto different axes. This organisation allows to perform a type of divide-and-conquer approach where we can at a first step explore the fundamental limitations of what is achievable within each category (or different combinations of these categories) as well as to explore connections with other well established disciplines in computer science and mathematics in general.

Here we briefly present the basics of the PRESTO framework~\cite{PZ}, which is an acronym for  Persistence, Robustness, Efficiency, STability, and Optimality (see Fig.~\ref{fig:spore}). We discuss each of these below, in reverse order:


\begin{figure}[t]
  \centering
    \includegraphics[width=0.34\textwidth, trim={4cm 2cm 4cm 1.8cm}, clip]{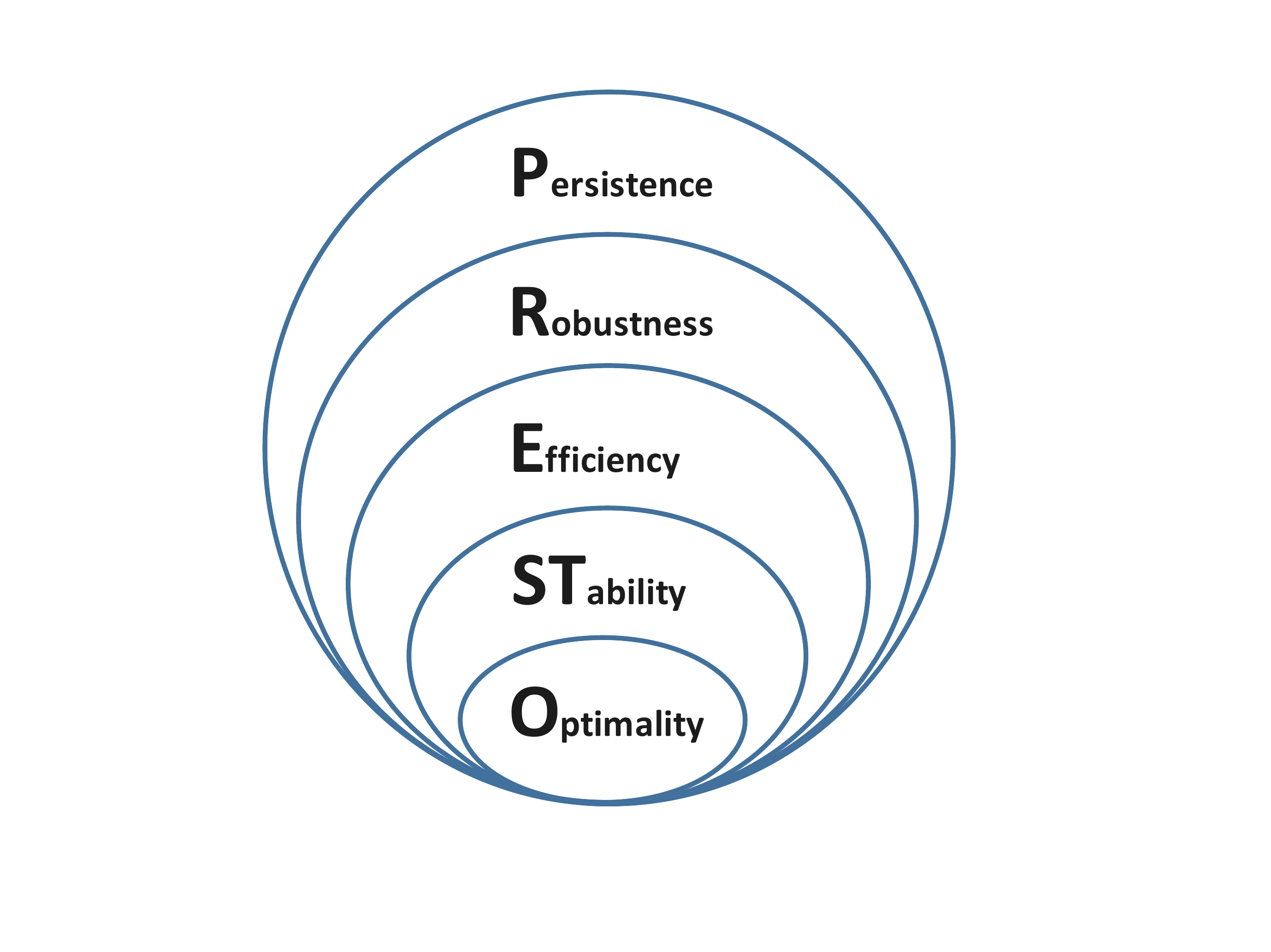}
      \caption{Graphical representation of the PRESTO framework.}
    \label{fig:spore}
\end{figure}

\subsubsection*{Optimality}

Optimality concerns the question of whether the protocol maximises the quality of certain outcomes. 
Ideally the protocol  should be mathematically proven to provide these guarantees.

Optimality is one of the most basic properties that a protocol can satisfy. It can be ascertained using mathematical analysis techniques that range from the very basic (e.g., calculus) to the more advanced (e.g., optimisation theory) \cite{gill1981practical,papadimitriou1998combinatorial}.
In the context of blockchain protocols, examples of optimality include formal proofs of liveness and safety under ``standard" network/agent conditions --- see, e.g.,\ \cite{cachin2017blockchains} for an overview of these properties for several recently implemented consensus protocols that are based on Byzantine fault tolerance.

\subsubsection*{Stability}

Stability means that it is in the participants' best interest to follow the rules of the protocol. In concrete terms, this means that the protocol is a Nash equilibrium (or more generally an incentive compatible outcome, e.g. a dominant strategy). An equilibrium is an outcome that is optimal from the perspective of all decision makers involved. 

This has been reasonably thoroughly studied (since 1940s, Nash \cite{nash}, von Neumann \cite{Neumann1944}). 
Game theory and economics provides numerous tools to model and analyze such settings. Importantly, 
stability does not imply optimality. The  ``price of anarchy" literature studies exactly this tension between equilibration and efficiency of outcomes \cite{KoutsoupiasP99WorstCE} and shows how in many natural settings equilibria although not perfectly optimal can be approximately optimal. Mechanism design on the other hand aims to re-design the rules of games so that individual incentives are perfectly aligned with societal goals. 
For blockchain protocols, stability means that it is best for agents in the default setting to follow the reference protocol.

\subsubsection*{Efficiency}

Does the protocol make efficient use of its computational resources? Does it perform its core tasks as fast as possible, using as little space as possible, using as little randomness as possible, using as little energy as possible, and using parallelisation efficiently? 

Efficiency of computation, at least from the perspective of time and space, has been reasonably thoroughly studied (since 1940s, Turing, von Neumann). Computational complexity theory provides the most carefully designed framework for studying the fundamental limitations of efficient computation \cite{papadimitriou2003computational}. These limitations force us to consider trade-offs, e.g., approximate optimality versus speed (e.g., approximation algorithms \cite{vazirani2013approximation}).  Ideally, a protocol implements a (near) optimal equilibrium efficiently.  Algorithmic game theory and mechanism design study questions on the intersection of optimality, efficiency and stability \cite{Nisan:2007:AGT:1296179}. For a blockchain example, the use of computational resources by the Bitcoin protocol is not efficient: the maximum transaction throughput is the same as five years ago despite a dramatic increase in hash rate and energy consumption \cite{odwyer2014bitcoin}.

\subsubsection*{Robustness}

Suppose that the protocol is reasonably close to an optimal efficient equilibrium on paper. Real distributed systems pose more challenges
(e.g. asynchrony, communication delays, users might have different utility functions due to differences in electricity/computation costs/risk attitudes, collusion, etc.). 

This is an emerging subfield within algorithmic game theory. A particular question of interest is what happens in the case of coalition formation or collusion. E.g., in \cite{eyal2014majority} it is shown that if $x > \frac{1}{3}$ fraction of the total mining power is owned by a mining pool, then the following the Bitcoin protocol is not an equilibrium. Specifically, the paper describes a strategy that can be used by a minority pool to obtain more revenue than the pool's fair share, i.e., more than its ratio of the total mining power. In \cite{sapirshtein2016optimal,kiayias2016blockchain} more elaborate attacking strategies are explored, as well as nearly-tight thresholds on the computational power of the attacker under which the honest strategy remains a Nash equilibrium.

\subsubsection*{Persistence}

Finally, what if the protocol is subjected to a severe attack or black-swan event, can the system recover? How fast, and at what cost?
For persistence, we take this idea to its logical extreme.
We assume that the system may be always under attack, and design it so
that it recovers and provides the desirable properties often.

Formally, we define a \emph{strongly persistent} property to be eventually be satisfied (and stay
satisfied) given any initial system condition.
By contrast, a \emph{weakly persistent} property will eventually be satisfied given any
initial system condition and will become satisfied again infinitely
often \cite{piliouras2014persistent}.
 Such ideas have been introduced within evolutionary game theory \cite{Hofbauer98} as well as the study of biological systems (i.e. recovery of a ecosystem after infection from a virus) \cite{smith2011dynamical}, however, the combination of these ideas with tools from optimisation theory and algorithm design have not been really explored.  \cite{piliouras2014persistent,piliouras2014optimization} perform some exploratory steps in this direction.

A desirable property is not satisfied by a system just in equilibrium,
but it is satisfied in a dynamic way.
This allows for more flexibility to explore trade-offs between recovery/convergence
time, `periodicity', and the cost of implementation.
Note that two (or more) incompatible properties can both be supported in a
weakly persistent manner.

\bigskip 
Summing up, the PRESTO framework sees protocols as a nesting doll with the following cascade of goals. First, optimality requires that the protocol solves the problem that it is defined to address, otherwise there is no good reason to deploy it and the designer should go back to the drawing board. Second, efficiency requires that resources are used as efficiently as possible (e.g. time, space, network bandwidth, energy, randomness, etc.). Next, stability analysis aims to make sure that self-interested agents have an incentive to follow and implement the protocol, i.e., that the protocol itself is an equilibrium. If not the agents will deviate from it and the deployed protocol will behave unpredictably in practice. Given an optimal, stable and efficient protocol we can then start considering more complicated behavioral models from the perspective of the agents. How robust is the equilibrium and its properties if we ``perturb'' the underlying assumptions, e.g., what if agents are risk-averse, or can form coalitions, how do the equilibria of these more realistic models compare to those idealised equilibria in the more vanilla settings? Finally, persistence goes beyond equilibrium thinking and asks which properties can be guaranteed not at equilibrium (i.e., consistently and at every single point in time), but maybe only recurrently (e.g., at consistent intervals, like once a week). This flexibility might allow us to provide guarantees that are impossible to satisfy in equilibrium. For example, two mutually exclusive properties cannot be both implemented at equilibrium, but can be guaranteed in a persistent manner by a system that cycles through different states. 

Exploring these trade-offs is a fascinating area for  multidisciplinary  research that could help us improve upon the understanding and design of the current protocols, as well help incorporate ideas from other well-established fields (e.g., algorithmic game theory, evolutionary game theory, complexity theory, approximation algorithms, a.o.).

\section{Conclusions and Recommendations }
\label{sec:conclusions}
In this paper, we have presented an overview of the security challenges in blockchain technology, and proposed ways to go forward. 
Our first step was to create the, to our knowledge, first structured attempt at cataloguing the wide range of blockchain-related security incidents. We find that apart from traditional OPSEC, many incidents are of a nature that is specific to blockchain technology. Two categories stand out: smart contracts, and techno-economic protocol incentives. In both settings, we argue that it is vital to properly test before deployment, because faults cannot be corrected without a hard fork. For the smart contracts, we have presented four testing approaches, building on the latest developments in automated code testing such as fuzzing, mutating, and model checking. For the protocol-level incentives, we have proposed a framework for comparing the security properties of protocols by ranking them across five dimensions. It is our position that for blockchain security to become mature, a rethink that leads to comprehensive testing before deployment is a necessity.


\bibliographystyle{unsrt}
\bibliography{blockchain_refs,darkweb_refs,refer}

\end{document}